\documentclass[twocolumn,superscriptaddress,amsmath,amssymb]{revtex4}
\usepackage{graphicx}
\usepackage{amssymb}

\newcommand{\nbaloxnb}{Nb/Al/\alox/Nb }
\newcommand{\alox}{AlO$_x$}
\newcommand{\kacm}{kA/cm$^2$}
\newcommand{\IV}{$I$-$V$}

\begin{document}

\title{Electrical stress effect on Josephson tunneling\\
through ultrathin AlO$_{x}$ barrier in \nbaloxnb junctions}

\author{Sergey K. Tolpygo}
\email{stolpygo@hypres.com}
\affiliation{HYPRES, Inc. 175 Clearbrook Road, Elmsford, NY 10523, USA}
\affiliation{Department of Physics and Astronomy, Stony Brook University\\
 Stony Brook, NY 11794-3800, USA}
\affiliation{Department of Electrical and Computer Engineering, Stony Brook University\\
 Stony Brook, NY 11794-2350, USA}

\author{Denis Amparo}
\email{denis.amparo@sunysb.edu}
\affiliation{Department of Physics and Astronomy, Stony Brook University\\
 Stony Brook, NY 11794-3800, USA}

\date{\today}

\begin{abstract}
The effect of dc electrical stress and breakdown on Josephson and quasiparticle tunneling in \nbaloxnb junctions with ultrathin \alox\ barriers typical for applications in superconductor digital electronics has been investigated. The junctions' conductance at room temperature and current-voltage (\IV) characteristics at 4.2 K have been measured after the consecutive stressing of the tunnel barrier at room temperature. Electrical stress was applied using current ramps with increasing amplitude ranging from 0 to $\sim1000I_\mathrm{c}$ corresponding to voltages across the barrier up to $\sim 0.65$ V, where $I_{\mathrm{c}}$ is the Josephson critical current. A very soft breakdown has been observed with polarity-dependent breakdown current (voltage). As the stressing  progresses, a dramatic increase in subgap conductance of the junctions,  the appearance of subharmonic current steps, and gradual increase in both the critical and the excess currents as well as a decrease in the normal-state resistance have been observed. The observed changes in superconducting tunneling suggest a model in which a progressively increasing number of defects and associated additional conduction channels (superconducting quantum point contacts (SQPCs)) are induced by electric field in the tunnel barrier. By comparing the \IV\ characteristics of these conduction channels with the nonstationary theory of current transport in SQPCs based on multiple Andreev reflections by Averin and Bardas, the typical transparency $D$ of the induced SQPCs was estimated as $D\sim0.7$. The number of induced SQPCs was found to grow with voltage across the barrier as $\sinh(V/V_0)$ with $V_0 = 0.045 $ V, in good agreement with the proposed model of defect formation by ion electromigration. The observed polarity dependence of the breakdown current (voltage) is also consistent with the model. Based on the observed magnitude of breakdown currents, electric breakdown of \alox barrier during plasma processing was considered to be an unlikely cause of fabrication-induced, circuit pattern-dependent nonuniformities of Josephson junctions' critical currents in superconductor integrated circuits.
\end{abstract}

\maketitle

\section{INTRODUCTION\label{sec:INTRODUCTION}}

Aluminum oxide (\alox) is widely used as a barrier material in various
applications involving tunnel junctions. It was also considered as
a potential gate oxide in advanced memory devices and metal-oxide-semiconductor
(MOS) transistors. Dielectric reliability issues such as oxide barrier
stability, leakage currents, and electric breakdown are very important
for electronics applications, especially for magnetic tunnel junctions
(MTJs) used for magnetic random access memories and superconducting
tunnel junctions (STJs) used for superconductor digital circuits requiring
ultrathin ($\sim1$ nm) tunnel barriers. The physics of dielectric
breakdown in ultrathin barriers is of great interest in its own right.
Oxide breakdowns are usually classified into two modes: intrinsic
and extrinsic. Though the difference is somewhat blurry, extrinsic
breakdowns are those caused by defects introduced or created during
oxide growth, whereas intrinsic ones are the property of a perfect
dielectric. In relatively thick oxide layers such as those used as
gate dielectric in MOS transistors, electric breakdown usually proceeds
by the accumulation of defects (traps) inside the dielectric until a percolation
pass is formed, at which point the resistivity suddenly decreases
from a very high value to a very low value, and a hard breakdown occurs.\cite{Degraeve1998,Chang2001}
In thinner oxide layers the breakdown often has a soft character which
is characterized by small gradual changes in resistance.\cite{Alam2002a,Alam2002b}
The thickness of ultrathin tunnel barriers used in superconductor
electronics is a couple of oxide monolayers, only a few interatomic
distances. Therefore, any defects formed in the oxide as a result
of electrical stress (e.g., displaced ions, oxygen vacancies, etc.)
should create additional conduction channels with significantly increased
transmission probability and consequently dramatically alter the quasiparticle
and Cooper-pair tunneling. In other words, the percolation path forming
at breakdown may consist  of just a single defect (trap) and hence
the breakdown may be very soft.

There have been several publications on the reliability and breakdown
of aluminum oxide layers with thicknesses above $\sim3$ nm as a new
gate oxide and $\sim1$ nm in MTJs.\cite{Shimazawa2000,Das2001,oliver2004,Kim2005,Akerman2006}
The existence of both intrinsic and extrinsic breakdown modes was
suggested.\cite{oliver2004} The intrinsic mode was associated with
a hard breakdown, and was suggested to be related to the chemical
bond breaking in applied electric field.\cite{Hill1983a,Hill1983b,McPherson1998}
The extrinsic mode was associated with a soft, gradual breakdown.
It was suggested to be related to pre-existing pinholes in the barrier
which grow in area as breakdown progresses due to Joule heating and/or
electric field effect.\cite{oliver2004}

Superconductor-insulator-superconductor (SIS) junctions offer unique
opportunities in studying breakdown mechanisms in ultrathin oxides
because both the quasiparticle and Josephson tunneling in STJs are
extremely sensitive to the barrier properties and boundary conditions
at the metal-oxide interfaces. In contrast to MTJs, pre-existing pinholes
in SIS junctions are easily identifiable because they carry supercurrent
thus creating nonuniform Josephson current distribution and dramatically
increasing subgap conductance. Whereas a microshort in MTJs was associated
with a junction having the resistance-area product $RA$ of $\sim$0.8
$\Omega\cdot\mu$m$^{2}$, and no tunneling magnetoresistance,\cite{oliver2004}
STJs with even lower values of $RA$ product exist and display interesting
Josephson tunneling properties.\cite{Kleinsasser1993,Miller1993,Patel1999}

Superconductor digital electronics utilizing SIS junctions has a potential
for sub-THz clock frequencies and ultra-low power dissipation for
digital signal processing, high-performance communications and computing.\cite{Likharev1991}
Recently, complex superconducting circuits based on Rapid Single Flux
Quantum (RSFQ) logic such as analog-to-digital converters and digital
RF receivers containing thousands of logic gates with clock frequencies
$\sim$30 GHz have been demonstrated, operating not only in liquid
He but also on commercial closed-cycle cryocoolers.\cite{Vernik2007}
Increasing the clock frequencies of superconductor integrated circuits
to $\sim$ 100 GHz would require employing high-$J_{\mathrm{c}}$
junctions with $RA$ products below $\sim$1 $\Omega\cdot\mu$m$^{2}$,
perhaps the thinnest tunnel barriers among all known devices.\cite{Chen1998,TolpygoASC2006}

Dielectric reliability may not appear to be important for superconducting
digital circuits because they operate at very low temperatures and
at very low voltages ($\sim$1 mV). Its significance however arises
from the possibility that tunnel barrier degradation may occur during
integrated circuit fabrication. For instance, the current state of
the art in RSFQ circuits has been plagued by limited circuit yield
brought about to a large extent by fabrication-induced variations
on the Josephson critical current ($I_{\mathrm{c}}$) of the tunnel
junctions.\cite{TolpygoASC2006,TolpygoISEC2007} These variations
may be related to dielectric barrier degradation and electrical breakdown.
It has been observed, for example, that the $I_{\mathrm{c}}$ of Josephson
tunnel junctions may depend on how the junction is wired to other
circuit elements, and in particular, at which step in the fabrication
process the junction makes electrical contact with the circuit's
ground plane.\cite{TolpygoISEC2007} For series arrays of nominally
identical tunnel junctions, it was frequently observed that the Josephson
critical current of a few junctions (usually of the first and the
last junction in the array) is significantly larger than for the rest
of the junctions. This cannot be simply explained by a variation in
the area of that one junction coming from the lithography and etch
processes of junction definition. Neither can it be explained by a
random fluctuation in the tunnel barrier transparency in that particular
junction, considering that the effect reproduces in different arrays
and the junctions in the array are just $\sim$10 $\mu$m apart. Instead,
it was suggested that the above phenomena are brought about by electrical
currents flowing through the tunnel barriers, a result of plasma processing
steps which follow the SIS trilayer deposition. Recent experiments
involving tunnel junctions protected from plasma process-induced electric
stress by current-limiting resistors support this suggestion.\cite{TolpygoEUCAS2007}

Surprisingly, there has been almost no research on the reliability
of ultrathin \alox\ barriers in STJs, except for early works on Al/\alox/Pb
junctions which studied the effects of electric annealing on the barrier
thickness, height and asymmetry of \alox\ barriers formed by plasma
oxidation.\cite{Konkin1980} These junctions, however, have no practical
application, and the changes in superconducting and Josephson properties
were not studied.

In this paper the effect of applied dc electrical stress on quasiparticle
and Josephson tunneling in \nbaloxnb junctions was investigated.
The study focused on this type of junctions because of their dominant
use in superconductor digital and analog electronics.

\section{FABRICATION\label{sec:FABRICATION}}

The \nbaloxnb junctions used in this study were fabricated at HYPRES,
Inc. using an 11-level process for superconductor integrated circuits.\cite{TolpygoASC2006,HypresDesignRules}
The fabrication was performed on 150-mm Si wafers. The process is
based on in-situ \nbaloxnb trilayer deposition.\cite{Gurvitch1983}
Specifically, Nb/Al bilayer (150 nm and 8 nm, respectively) deposition
is followed by \alox\ formation by room temperature oxidation of Al.
The \alox\ layer is then topped off by the deposition of the Nb counter-electrode
(50 nm). All metal layers are deposited by dc magnetron sputtering
in a cryopumped vacuum system with base pressure of $1\times10^{-9}$
Torr. Two Josephson critical current densities, $J_{\mathrm{c}}$
(1 \kacm\ and 4.5 \kacm), were targeted, obtained by Al oxidation
for 15 min at oxygen pressure of 170 mTorr and 18 mTorr, respectively.
After the counter-electrode etch process that defines the junctions, their interior was sealed
along the perimeter and sidewalls by an anodization layer composed
of Al$_{2}$O$_{3}$ and Nb$_{2}$O$_{5}$ in order to protect the
barrier from reacting with process chemicals used in subsequent fabrication
steps.

Circular JJs with design radii of 2.00 $\mu$m ($A=12.6$ $\mu$m$^{2}$)
and 0.95 $\mu$m ($A=2.8$ $\mu$m$^{2}$) for wafers with $J_{\mathrm{c}}=1$
\kacm\ and 4.5 \kacm\ were arranged on $5\times5$-mm$^{2}$ chips
referenced according to wafer number and the coordinates (in units
of 5 mm) of the chip location on the wafer. For example, a chip from
wafer KL1004 with location (-5,7) is to be called KL1004N5P7.

\section{EXPERIMENT\label{sec:EXPERIMENT}}

Each chip was mounted inside a magnetically shielded cryoprobe. The
junctions were measured using a low-pass filtered four-probe setup
with a Keithley 2000 voltmeter and a Keithley 6220 current source.
Electrical stress was applied at room temperature by ramping current
through the junction up to a preselected value $I_{\mathrm{S}}$ and
back down to zero. The stress application was preceded and followed
by a measurement of the junction's room-temperature resistance, using
a low current of 500 $\mu$A. The tunneling \IV\ characteristics were
then measured in the superconducting state at $T=4.2$ K with the
cryoprobe submerged in a liquid He. The next stress/measurement cycles
were then performed using progressively higher values of $I_{\mathrm{S}}$.

The effect of stress polarity was also investigated. Here we define
positive stress as current flowing from the counter electrode (positive
potential on the top Nb layer) to the base electrode (Nb/Al bilayer),
while negative stress is current flowing in the opposite direction.
At room temperature, there is a resistance in series with the tunnel
barrier associated with the normal resistance of interconnects to
the junction. This series resistance was estimated from the interconnects'
geometry using a separately measured sheet resistance of the layers
involved. This resistance was assumed to remain unchanged by stress
applications and was simply subtracted from the total measured resistance,
thus allowing us to estimate the potential difference that develops
across the tunnel barrier during stressing. Parameters of some of
the studied junctions are given in Table \ref{tab:sample-parameters}.

\section{EXPERIMENTAL RESULTS\label{sec:EXPERIMENTAL-RESULTS}}

\begin{figure}
\includegraphics{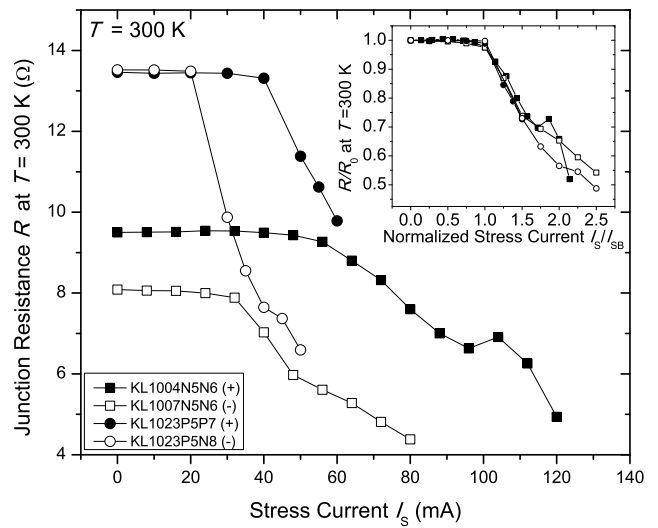}

\caption{\label{fig:RTResistanceVsStressCurrent} Junction resistance at room temperature
after electrical stressing of the junctions in Table \ref{tab:sample-parameters};
(+)/(-) indicates positive/negative stress polarity. Despite small
differences between individual junctions, the stress-induced irreversible
resistance changes are very similar as demonstrated in the Inset showing
the normalized resistance $R/R_{\mathrm{n}}$ at room temperature as a function of normalized
stress current $I_{\mathrm{S}}/I_{\mathrm{SB}}$ for the same junctions
(see Table \ref{tab:sample-parameters}).}

\end{figure}

Fig. \ref{fig:RTResistanceVsStressCurrent} shows the resistance at
low currents ($ < 500$ $\mu$A) of four \nbaloxnb junctions after each
subsequent stress application. The junction resistance remains roughly
constant after stressing with low currents until a threshold stress
current for breakdown $I_{\mathrm{SB}}$ is reached, above which the
resistance starts to decrease indicating irreversible changes in the
barrier. Each succeeding decrease in resistance is apparently a cumulative
effect of all previous stress applications. The threshold current
varies among nominally identical junctions on the wafer and from wafer
to wafer, suggesting some statistical nature of the barrier breakdown.
Despite these variations, the observed threshold current for the positive
stress (current from Nb counter-electrode to Al) is consistently higher than for
negative stress as listed in Table \ref{tab:sample-parameters}. If the
resistance after stressing is scaled with the resistance in the initial
junction and the stress current is scaled with the threshold current,
all the curves in Fig. \ref{fig:RTResistanceVsStressCurrent} collapse
onto a single curve (see Fig. \ref{fig:RTResistanceVsStressCurrent}
inset) suggesting a universal breakdown mechanism and a universal
character of resistance changes due to electrical stress.

The irreversible decrease of the tunnel barrier resistance may be
due to (a) a decrease in the average barrier height and/or thickness;
(b) the barrier becoming nonuniform due to formation of additional
conduction channels (regions with increased barrier transparency which
are often called micro- or nano-shorts); and (c) a combination of
the above. However, room temperature measurements alone are insufficient
to distinguish between these possibilities and measurements in the
superconducting state of the junction electrodes are needed.

\begin{figure}
\includegraphics{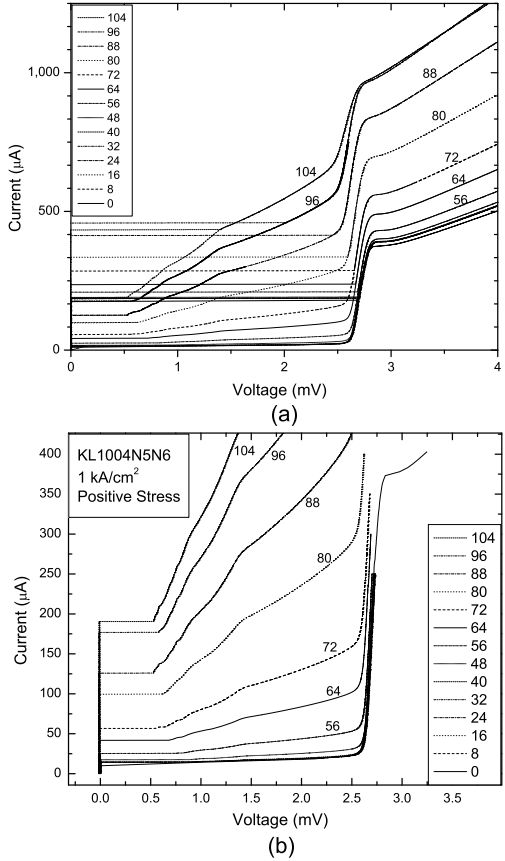}

\caption{\label{fig:KL1004N5N6-IVCurves} (a) \IV\ characteristics of \nbaloxnb
junction at $T=4.2$ K (initial $J_{\mathrm{c}}=1$ \kacm) after
each application of electrical stress show increasing $I_{\mathrm{c}}$,
decreasing $R_{\mathrm{n}}$, increasing subgap conductance, and increasing
excess current. The gap voltage $V_{\mathrm{g}}=2\Delta/e$ and the
current step at $V=V_{\mathrm{g}}$ remain almost unaffected by electric
stress in the wide range of stress currents from 0 up to $\sim2I_{\mathrm{SB}}$.
At higher stress currents the gap structure broadens and diminishes,
and at $I_{\mathrm{S}}\sim3I_{\mathrm{SB}}$ the junction loses all
remaining signatures of the tunnel junction. Numbers in the legend indicate the applied positive stress current in mA and identify the curves from top to bottom. (b)
Blow-up of the return branches of \IV\ curves of KL1004N5N6 after
each stress application clearly shows the development of current steps
(subgap structure) at subharmonics of the gap voltage.}
\end{figure}

The electric stress-induced changes in the tunnel barrier properties
are clearly seen in the Josephson \IV\ characteristics at $T=4.2$K shown
in Fig. \ref{fig:KL1004N5N6-IVCurves}(a). Five main features are worth
mentioning. First, the Josephson critical current, $I_{\mathrm{c}}$
(defined here as the switching current) increases with the stress
current. Second, the normal-state resistance, $R_{\mathrm{n}}$ of
the junction decreases with the stress current. Third, the so-called
knee structure in the quasiparticle tunneling just above the gap voltage
of the junction also shifts to higher currents as the stress increases.
Fourth, the junction conductance in the subgap region of voltages
increases with the stress current. The retrapping current also increases
with the increase in subgap conductance. Finally, the gap voltage
$V_{\mathrm{g}}$ is nearly independent of the stress current and
decreases only slightly at very high stress currents. The increase
in the subgap conductance is by far the most pronounced change in
the \IV\ curves of the stressed junctions. For instance, after applying
a stress current of 96 mA, the Josephson critical current of the junction
increases by a factor of $\sim2.5$ whereas the subgap conductance
at 2 mV increases by a factor of $\sim25$.

\begin{table*}
\caption{\label{tab:sample-parameters}Summary of initial resistances, initial critical current, and  breakdown stress current for the samples shown in Fig. \ref{fig:RTResistanceVsStressCurrent}}
\begin{ruledtabular}
\begin{tabular}{ccccccc}
 & Target $J_{\mathrm{c}}$  & Stress  & $I_{\mathrm{SB}}$  & $R_{0}$ at 300 K & $R_{\mathrm{n}0}$ at 4.2 K & $I_{\mathrm{c}0}$ at 4.2 K\\
 & (\kacm)  & Polarity  & (mA)  & ($\Omega)$  & ($\Omega)$  & ($\mu$A)\\
\hline 
KL1004N5N6  & 1.0  & +  & 56  & 9.27  & 7.22  & 180\\
\hline 
KL1007N5N6  & 1.0  & -  & 32  & 7.88  & 6.21  & 177\\
\hline 
KL1023P5P7  & 4.5  & +  & 40  & 13.31  & 10.84  & 118\\
\hline 
KL1023P5N8  & 4.5  & -  & 20  & 13.48  & 10.67  & 120\\
\end{tabular}
\end{ruledtabular}
\end{table*}

\begin{figure}
\includegraphics{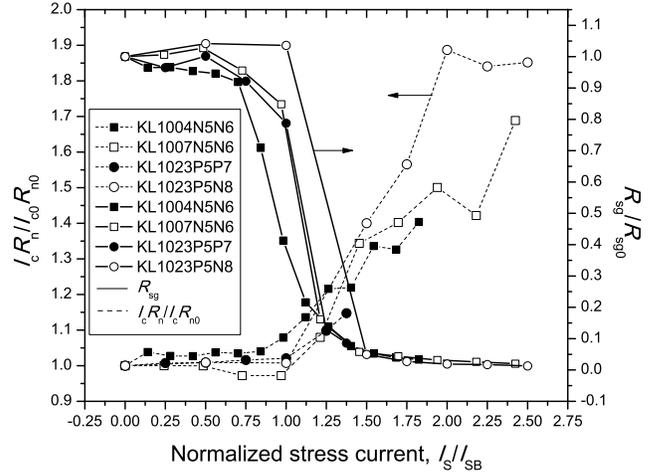}

\caption{\label{fig:icrn-rsg} $I_{\mathrm{c}}R_{\mathrm{n}}$ product (left
scale, dotted lines) and $R_{\mathrm{sg}}$ at 2 mV (right scale,
solid lines) values after each stress application normalized to their
initial values $I_{\mathrm{c0}}R_{\mathrm{n}0}$ and $R_{\mathrm{sg}0}$
in the unstressed junctions. Parameters of the junctions are given
in Table \ref{tab:sample-parameters}.}

\end{figure}

\begin{figure}
\includegraphics{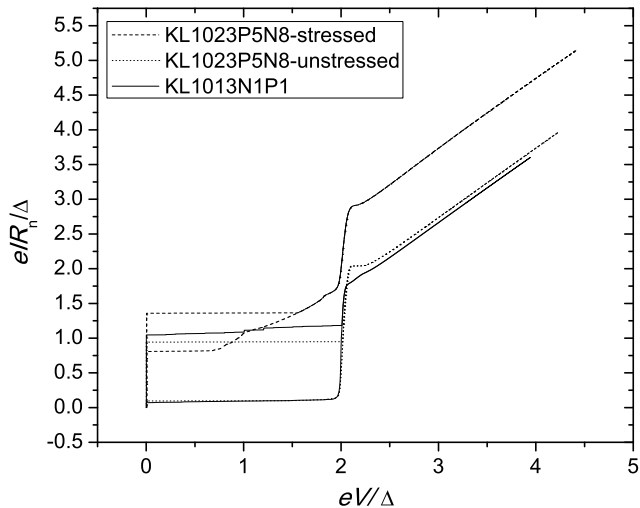}

\caption{\label{fig:stressed-unstressed}A comparison of the \IV\ curve of
a stressed junction KL1023N5P8 (dashed curve) with post-stress Josephson critical
current density $J_{\mathrm{c}}\sim9$ \kacm\ (initial $J_{\mathrm{c}}=4.5$ \kacm)
and an as-fabricated, unstressed junction KL1013N1P1 (solid curve) with $J_{\mathrm{c}}\sim11$ \kacm.
Although the as-fabricated junction has an even larger $J_{\mathrm{c}}$ (larger
average barrier transparency), its \IV\ curve is very different from
the electrically stressed junction: it has no appreciable subgap conductance,
no subharmonic current steps, and no excess current; its $I_{\mathrm{c}}R_{\mathrm{n}}$
product and the current step at $V_{\mathrm{g}}$ are close to the values given
by the microscopic theory for tunnel junctions with low transparency.
Presumably, the as-fabricated junctions have a uniform tunnel barrier
whereas the barrier in electrically stressed junctions becomes
nonuniform. The \IV\ curve of the initial, unstressed junction KL1023N5P8 is also shown (dotted curve).}
\end{figure}

\begin{figure}
\includegraphics{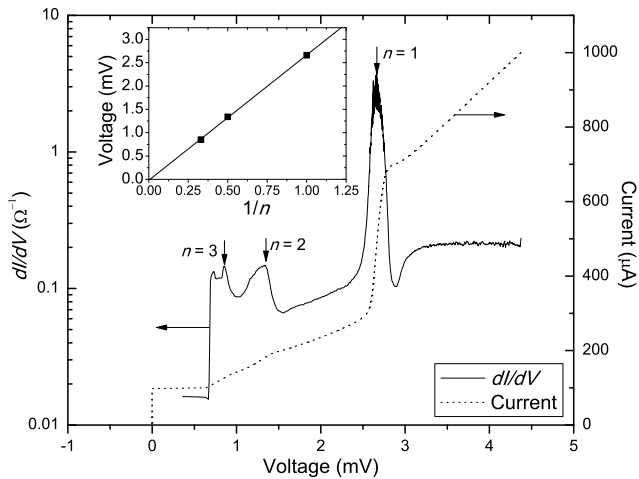}

\caption{\label{fig:mar-peaks}The return branch of the \IV\ characteristic (dotted curve)
of KL1004N5N6 after application of 80 mA stress along with differential
conductance $dI/dV$ (solid curve) showing peaks corresponding to multiple Andreev
reflections (MAR) of quasiparticles. Inset shows the voltages corresponding to conductance
peaks; the straight line is a fit to $2\Delta/en$ ($n=1,2,\ldots$) dependence expected for MAR,
giving $2\Delta/e = 2.696$ mV.}

\end{figure}

\section{DISCUSSION\label{sec:DISCUSSION}}

The experimental results presented in the previous section show that
there is a broad range of electric stress currents through \nbaloxnb
junctions (and corresponding voltages across the \alox\ tunnel barrier)
where a gradual change of the barrier properties occurs. This can
be interpreted as a soft breakdown of the oxide barrier in contrast
to the sudden changes (hard breakdown) usually observed in thicker
oxide layers of SiO$_{2}$ and \alox. For the junctions with $J_{\mathrm{c}}=1$
\kacm, this range spans from the typical threshold current $I_{\mathrm{SB}}\sim30$
mA (corresponding to a stress current density $J_{\mathrm{S}}\sim2.4\times10^{5}$
A/cm$^{2}$) to $I_{\mathrm{S}}\sim120$ mA ($J_{\mathrm{S}}\sim10^{6}$ A/cm$^{2}$)
above which the stressed devices lose all the signatures of superconducting
tunnel junctions. For the positive stress polarity, the threshold
current in Table \ref{tab:sample-parameters}, $I_{\mathrm{SB}}=56$ mA corresponds
to the voltage across the barrier $V_{\mathrm{b}}^{+}=$ 0.519 V. For the negative
stress polarity, $I_{\mathrm{SB}}=32$ mA corresponds to $V_{\mathrm{b}}^{-}=0.252$
V. The typical barrier thickness $d$ for the Josephson current densities
used in this work is $\sim1$ nm or less.\cite{Tolpygo2003,Cimpoiasu2004}
Hence, the typical electric fields across the \alox\ barrier at which
the irreversible changes start are $E_{\mathrm{b}}^{-}=$ $2.52\times10^{8}$
V/m and $E_{\mathrm{b}}^{+}=5.19\times10^{8}$ V/m for the negative and positive
stress polarities, respectively. 

The difference between $V_{\mathrm{b}}^{+}$ and $V_{\mathrm{b}}^{-}$ was also observed in 
\alox\ breakdown measurements in magnetic tunnel junctions\cite{Kim2005,Akerman2006} where it was
speculated to be a result of different surface roughness of the two
metal/oxide interfaces. We suggest that the difference between $V_{\mathrm{b}}^{+}$
and $V_{\mathrm{b}}^{-}$ is more fundamental and is a result of difference
in work functions of the junction electrodes. Indeed, from the contact
potential difference measurements, the work functions in the electrodes
are $\varphi_{\mathrm{Al}}=4.19$ eV and $\varphi_{\mathrm{Nb}}=4.37$
eV for Al and Nb, respectively. Therefore, in a \nbaloxnb junction
the potential barrier becomes asymmetric and there is an internal
electric field across the barrier $E_{\mathrm{int}}=\Delta\varphi_{\mathrm{Nb/Al}}/ed\sim$$1.8\times10^{8}$
V/m directed from the Nb/Al base electrode towards the Nb counter
electrode and arising from the difference in work function $\Delta\varphi_{Nb/Al}=\varphi_{Nb}-\varphi_{Al}=0.18$
eV. This barrier asymmetry agrees well with the result obtained from
the asymmetry of conductance vs. voltage characteristics of \nbaloxnb
junctions with low tunnel barrier transparency.\cite{Tolpygo2003,Cimpoiasu2004}
For the negative stress polarity (higher potential on Nb counter-electrode),
the external field adds up to the internal field $E_{\mathrm{t}}^{-}=E_{\mathrm{ext}}+E_{\mathrm{int}}$
in the same direction, whereas for the positive stress polarity the
external field is opposite to the internal one $E_{\mathrm{t}}^{+}=E_{\mathrm{ext}}-E_{\mathrm{int}}$,
where $E_{\mathrm{t}}$ is the total field in the dielectric and $E_{\mathrm{ext}}$
is the externally applied electric field. If defects in the barrier
start to form when the net internal field reaches some critical level
$E_{\mathrm{c}}$ (e.g., as a result of ion electromigration), from the onset
of irreversible resistance changes (soft breakdown), we estimate the
critical voltage to be $V_{\mathrm{c}}=(V_{\mathrm{b}}^{+}+V_{\mathrm{b}}^{-})/2=0.385$ V, corresponding
to $E_{\mathrm{c}} \approx 3.85\times10^{8}$ V/m.
The barrier asymmetry in this model corresponds to $(V_{\mathrm{b}}^{+}-V_{\mathrm{b}}^{-})/2 = 0.13$ V, in excellent agreement with the 0.18 V difference between work functions.

Let us see what can be inferred about the nature of stress-induced
changes to the barrier from the data presented in Sec. \ref{sec:EXPERIMENTAL-RESULTS}.
Clearly the increase of the critical current and decrease in the normal
resistance in stressed junctions indicate that the critical current
density and the average barrier transmission increase. The question
is whether the tunnel barrier remains uniform or not. If the barrier
remains uniform and only its height and/or thickness is changing with the
applied stress, one would expect the junction $I_{\mathrm{c}}R_{\mathrm{n}}$ product
to remain constant because in the limit of low
barrier transmission probability, it is given by  the Ambegaokar-Baratoff
(AB) relationship $I_{\mathrm{c}}R_{\mathrm{n}}=\pi\Delta/2e$ for $T\ll T_{\mathrm{c}}$,\cite{Ambegaokar1963-1}
and the gap voltage was found to remain nearly constant after barrier
stressing. Contrary to this, the measured $I_{\mathrm{c}}R_{\mathrm{n}}$ product increases
significantly in the stressed junctions as can be seen in Fig. \ref{fig:icrn-rsg}.
This may indicate a transition from tunnel junction to point contact
behavior because for point contacts between two superconductors
at $T\ll T_{\mathrm{c}}$ the $I_{\mathrm{c}}R_{\mathrm{n}}$ product is greater than in the
tunnel junction by a factor of 1.32 in the dirty limit and by a factor
of 2 in the clean limit, as was shown by Kulik and Omel'yanchuk (KO).\cite{Kulik1975,Kulik1978}
It was also found by Arnold\cite{Arnold1985,Arnold1987} that, for
an insulating and structureless barrier of arbitrary thickness, the
$I_{\mathrm{c}}R_{\mathrm{n}}$ product in general depends on the barrier transmission
probability $D$, and reduces to the AB result at $D\ll1$ and to
the KO result for the clean point contact at $D=1$.

It is easy to show, however, that the barrier does not remain uniform
and structureless after stressing. For this, we compare the \IV\ curves
of a stressed junction and an as-prepared, unstressed junction with
the same average Josephson critical current density. The as-prepared
junction is presumably as uniform as possible for the given fabrication
method. The \IV\ curves are shown in Fig. \ref{fig:stressed-unstressed}.
The stressed junction with $J_{\mathrm{c}}\sim9.3$ \kacm\ clearly has different
\IV\ characteristics (especially at $V<V_{\mathrm{g}}$) than the as-prepared
junction with an even higher $J_{\mathrm{c}}\sim11$ \kacm. This comparison
indicates that the changes in the \IV\ curves and the increase in
the $I_{\mathrm{c}}R_{\mathrm{n}}$ product brought about by electric stress cannot
be explained by a gradual increase in transparency of a uniform and
structureless barrier. A dramatic increase in the subgap conduction
in the stressed junctions, the appearance of pronounced features at
subharmonics of the gap voltage, $2\Delta/en$ at $n=2,3,\ldots$ shown
in Fig. \ref{fig:mar-peaks} along with the increase in $I_{\mathrm{c}}R_{\mathrm{n}}$ strongly
indicate that additional conduction channels with increased transparency
are gradually formed as a result of electric stress. These conduction
channels can be viewed as point contacts between the junction electrodes,
pinholes, or nanoshorts in the barrier, contributing to both
the quasiparticle and Cooper-pair transport in parallel to the main
tunnel barrier which remains largely unmodified by the stress.

Subgap features similar to those appearing in electrically stressed junctions (Fig. \ref{fig:KL1004N5N6-IVCurves}(b) and Fig. \ref{fig:mar-peaks}) have long been observed in \nbaloxnb and other types of junctions \cite{Kleinsasser1993,Miller1993,Patel1999} with high critical current densities and attributed to MAR. Two models were proposed to explain the appearance of MAR in high-$J_{\mathrm{c}}$ junctions. Kleinsasser et al. \cite{kleinsasser1994} suggested that, as oxygen exposure during Al oxidation decreases, the barrier becomes nonuniform, consisting of regions with low transparency (good tunnel barrier) and regions with high transparency (pin holes). That is, the transparency distribution has two sharp peaks, one at a low $D$ value and another one at $D\sim1$. As oxygen exposure decreases, the relative contribution of the second peak increases, and so does the $J_{\mathrm{c}}$. Naveh et al. \cite{Naveh2000} argued that the defects in the ultrathin tunnel barrier are naturally occurring and, therefore, the transparency distribution in high-$J_{\mathrm{c}}$ junctions is universal and given by the Schep-Bauer (SB) distribution\cite{Naveh2000} 
\begin{equation}
\rho(D) = \frac{G}{\pi G_0} \frac{1}{D^{3/2}\sqrt{1-D}}
\label{eq:schep-bauer}\end{equation}
They found that SB distribution gives a better fit to experimental \IV\ curves than averaging over the Dorokhov distribution describing the transparency distribution in long disordered conductors---not a surprise though because the Dorokhov distribution should not be applicable to short channels by definition. There is ample experimental evidence however that the behavior of high-$J_{\mathrm{c}}$ junctions is not universal. For instance, the value of $J_{\mathrm{c}}$ at which MAR steps become visible strongly depends on the junction fabrication procedure and varies between different experimental groups.\cite{Miller1993,Patel1999} Our experimental results show that MAR steps can be created at will in tunnel barriers with any initial transparency as a result of electrical stress. Therefore, what were perceived as natural and universal defects appearing in the oxide barrier as a result of a short oxidation process \cite{Mallison1995} could be simply a result of damage to the very thin tunnel barrier induced during the junction (wafer) fabrication processes. This could explain why improvements in the junction fabrication process usually shift the $J_{\mathrm{c}}$ level at which MAR steps begin to appear towards higher values. 

Although no microstructural characterization of the barrier changes
was undertaken in this work, a plausible scenario of how the additional conduction
channels are induced by electric stress can be proposed. It
is known that \alox\ formed by room temperature oxidation is amorphous
and nonstoichiometric, i.e. contains a large amount of oxygen vacancies.\cite{Tan2005}
At the oxidation conditions used in our work, the barrier thickness
is only $\sim1$ nm, i.e. just about two to three nearest-neighbor distances. Therefore,
a displacement of even a single atom from the barrier can significantly
increase local transparency. If a dc electric field is applied to
the oxide barrier, electromigration of cations (Al$^{3+}$) and anions
(O$^{2-}$, O$^{-}$) in opposite directions may occur. From anodic
oxidation of Al in electrolytes it is known that the anodizing ratio
for Al is $ar=1.3$ nm/V, i.e., 1.3 nm of \alox\ is formed for each
volt applied to the Al/electrolyte interface.\cite{Diggle1969} That
is, on average, oxygen ions are transported inside the aluminum electrode
by $\sim1$ nm per 1 volt of applied voltage. Ionic current density
in the oxide is a strong function of the electric field strength $J_{i}=\alpha\exp(\beta E-\gamma E^{2})$.\cite{Diggle1969,Dignam1968}
In our case, there is no electrolyte providing a constant supply of
oxygen atoms for the oxide to grow. However, some oxygen ions from
\alox\ barrier will electromigrate into Al underlayer of the Nb/Al/\alox\
structure if the Nb/Al base electrode is at a positive potential (negative
stress polarity in our nomenclature) while Al$^{3+}$ cations should
migrate towards the interface with Nb counter electrode. At a positive
potential on Nb counter electrode (positive stress), oxygen ions from
the \alox\ barrier will move into Nb counter electrode, depleting the
barrier. The average oxygen displacement at the onset of breakdown
can be estimated using the typical threshold voltage $V_{\mathrm{b}}^{-}=0.25$
V and $ar=1.3$ nm/V, giving $\sim0.3$ nm. This closely corresponds
with O-O nearest-neighbor distance of 2.8 \AA\ in amorphous alumina.\cite{Lamparter1968}

Any ion displacement in the barrier under the effect of an applied
electric field creates a defect and alters the barrier transparency.
The length of the newly created conduction channel is about the same
as the barrier thickness which is much less than the coherence length
$\xi$ and the magnetic field penetration depth $\lambda$. The cross-section
of such a channel is also on the atomic scale, much less than $\xi$
and $\lambda$. The theory of current transport through such short
and narrow channels, often called quantum point contacts (QPCs), is
well developed. For the single mode QPC, the Josephson current is given by

\begin{equation}
I=\frac{e\Delta}{2\hbar}\frac{D\sin\phi}{\sqrt{1-D\sin^{2}(\phi/2)}}\tanh\frac{\Delta\sqrt{1-D\sin^{2}(\phi/2)}}{k_{\mathrm{B}}T}\label{eq:josephson-current}\end{equation}
 where $D$ is the channel transparency and $\phi$ is the phase difference.\cite{Beenakker1991}

The theory of nonstationary properties in SQPCs was developed by Averin
and Bardas.\cite{Averin1995} It is based on the idea of multiple Andreev reflections
(MAR) \cite{Andreev1964} which reveal themselves as current steps
at subharmonics of the gap voltage $2\Delta/en$, at $n=2,3,\ldots$. They found that \IV\ characteristics
strongly depend on the SQPC transparency and demonstrate the most
pronounced subharmonic steps at intermediate values of $D\sim0.4$
to 0.5. The theoretical \IV\ curves reduce to a pure tunneling characteristics
at $D<0.1$ and to a featureless curve typical for an \IV\ of a clean
S-N interface at $D\sim1$.\cite{Averin1995}

\begin{figure}
\includegraphics{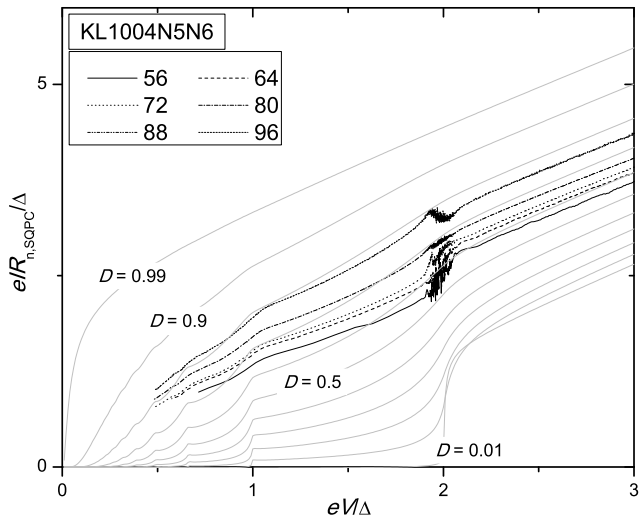}

\caption{\label{fig:averin-comparison}Normalized \IV\ characteristics of the additional conduction channels (SQPCs) created by positive electric stressing at currents $I_{\mathrm{S}} = 56,64,\ldots,96 $ mA. The curves were obtained by subtracting the \IV\ characteristics of the initial, unstressed junction $I_{0}(V)$ from the \IV\ curve after each stress application, $I_{\mathrm{SQPC}}(V) = I(V)-I_{0}(V)$. The dip at $eV/\Delta=2$ in the curves is an artifact of the subtraction procedure due to some broadening and slight decrease of the gap after electric stressing. Theoretical I-V curves for a single SQPC with varying transparency $D=0.01,0.1,0.2,0.3,\ldots,0.8,0.9,0.99$ (bottom to top) from \cite{Averin1995} are also shown. As can be seen, all the obtained experimental dependences fall within the range of theoretical curves corresponding to  $0.6 \lesssim D \lesssim 0.8$.}

\end{figure}

Let us see what properties of conduction channels created in the \alox
barrier by electrical stressing can be inferred from the measured
\IV\ curves shown in Fig. \ref{fig:KL1004N5N6-IVCurves}. Because
of space constraints we will restrict our analysis to the positive
stress polarity only. The case of the negative stress polarity is
completely analogous and will be presented elsewhere. First, we note
that the amplitude of the current step $\Delta I_{\mathrm{ss}}$ at $V_{\mathrm{g}}=2\Delta/e$
practically does not change in stressed junctions although the subgap
conductance dramatically increases. (In the first approximation we
will neglect some broadening of the current step at $V_{\mathrm{g}}$ and a
small decrease in $V_{\mathrm{g}}$ at extremely high stress currents which
could result from the smearing of the metal-insulator interfaces.)
In the microscopic theory of tunneling in superconductors, the size
of the current step is proportional to the area of the tunnel barrier
and is given by

\begin{equation}
\Delta I_{\mathrm{ss}}=G_{\mathrm{NN}}A\frac{\pi\Delta}{2e}\label{eq:current-step}\end{equation}
 where $G_{\mathrm{NN}}$ is the specific normal-state tunneling conductance.

\begin{figure}
\includegraphics{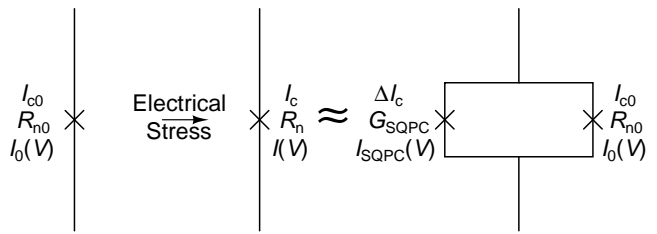}

\caption{\label{fig:sqpc-circuit-diagram}Circuit diagram of the proposed model.
In this model, the the application of electrical stress results in the formation
of few additional conduction channels $G_{\mathrm{SQPC}}$ in the tunnel barrier
which remains largely unchanged and is assumed to be the same as the initial,
unstressed junction.}
\end{figure}

It is easy to verify that the experimentally observed $\Delta I_{\mathrm{ss}}$
in Fig. \ref{fig:KL1004N5N6-IVCurves} is very close to the value given by Eq. (\ref{eq:current-step}).
The independence of $\Delta I_{\mathrm{ss}}$ of electric stress indicates
that the area of the tunnel barrier and its conductance does not change
noticeably. It means that the total area of additional conduction
channels formed by stressing is much smaller than the junction area.
Therefore, a stressed junction can, in the first approximation, be
represented by a parallel combination of the initial (unstressed)
junction and some number of SQPCs (see Fig. \ref{fig:sqpc-circuit-diagram}):
$I(V)=I_{\mathrm{0}}(V)+I_{\mathrm{SQPC}}(V)$. To obtain the \IV
characteristics of these SQPCs, $I_{\mathrm{SQPC}}(V)$ we can subtract
the \IV\ curve of the tunnel (unstressed) junction $I_{0}(V)$ from
the experimental $I(V)$ in the stressed junction. The result of this
procedure is shown in Fig. \ref{fig:averin-comparison}, along
with the theoretical \IV\ curves calculated for a SQPC with different
transparencies.\cite{Averin1995}

Although the subtraction procedure is not very accurate near the gap
voltage because of the gap smearing in the highly stressed junction,
it is clear that the \IV\ curve of the SQPCs has no significant current
step at $V=2\Delta/e$. This is true for any stress current $I_{\mathrm{S}}$ (any
number of SQPCs) because in all the \IV\ curves in Fig. \ref{fig:KL1004N5N6-IVCurves} the current step at $V=2\Delta/e$
is the same as in the initial junction. In the microscopic theory
of nonstationary properties of a single SQPC,\cite{Averin1995} the
size and broadening of the current step at $V=2\Delta/e$ in the \IV
curve as well as the size and broadening of subharmonic current steps
strongly depends on the channel transparency. The step at $2\Delta/e$
basically disappears at $D>0.6$ while subharmonic steps basically
disappear at $D>0.9$. Therefore, in the theory, the range of channel
transparencies where the \IV\ curves would look like in Fig.
\ref{fig:KL1004N5N6-IVCurves}(b) and Fig. \ref{fig:averin-comparison} is very
narrow $0.6\lesssim D\lesssim0.8$. Hence, it would not make a significant
error if, for simplicity, we assume that all the channels have the
same transparency. A better approach would only be a full fitting
of the \IV\ curves to the theory, which will be done elsewhere. It
can be seen from Fig. \ref{fig:averin-comparison} that all the
\IV\ curves of SQPCs formed in the stress current range from the breakdown
$I_{\mathrm{SB}}$ to 2$I_{\mathrm{SB}}$ lie between the theoretical
curves with $D=0.6$ and $D=0.8$. Therefore, the average channel
transparency formed at electric stressing can be taken as $\left\langle D\right\rangle =0.7$.

\begin{figure}
\includegraphics{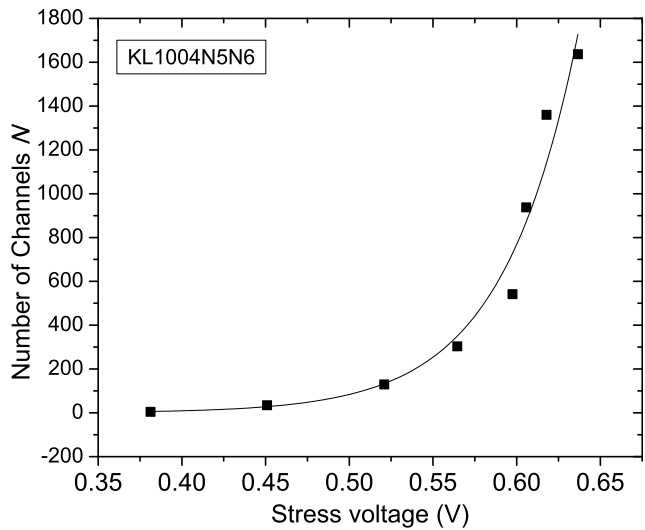}

\caption{\label{fig:sqpc-fit-NvsV}The number of created channels $N$ (assuming $\left\langle D\right\rangle =0.7$) versus
the maximum applied stress voltage. The solid curve is the fit to Eq. (\ref{eq:sinh})
yielding $V_{0}=0.045$ V. This is very close to the calculated value
of $V_{0}=0.046$ V.}

\end{figure}

From the linear part of the \IV\ curves at $V>V_{\mathrm{g}}$, the normal-state
conductance of the channels is simply $G_{\mathrm{SQPC}}=R_{\mathrm{n}}^{-1}-R_{\mathrm{n}0}^{-1}$
, where $R_{\mathrm{n}0}$ is the normal-state tunnel resistance in the initial
junction measured before stressing. Since the normal-state conductance
of a single QPC is given by $2e^{2}D/h$, we can estimate the number $N$ of created SQPCs using
$N\cdot\left\langle D\right\rangle $ = $G_{\mathrm{SQPC}}h/2e^{2}$. Its dependence
on the stress voltage is shown in Fig. \ref{fig:sqpc-fit-NvsV}.
At $\left\langle D\right\rangle =0.7$, the number of channels formed
at the maximum stress currents studied is $\sim2\times10^{3}$, corresponding
to an average channel density of $1.6\times10^{10}$ channels/cm$^{2}$
and an average channel spacing of $\sim80$ nm. In our model each
channel is associated with a displaced ion in the barrier, so the
estimates above give the surface density and the average distance
between displaced ions in the barrier at the maximum stress currents
$I_{\mathrm{S}}= 100$ mA. The obtained estimates are self-consistent with the
initial supposition that the number of additional conduction channels
is small and that the area occupied by the channel is negligible with
respect to the area of the tunnel barrier.

\begin{figure}
\includegraphics{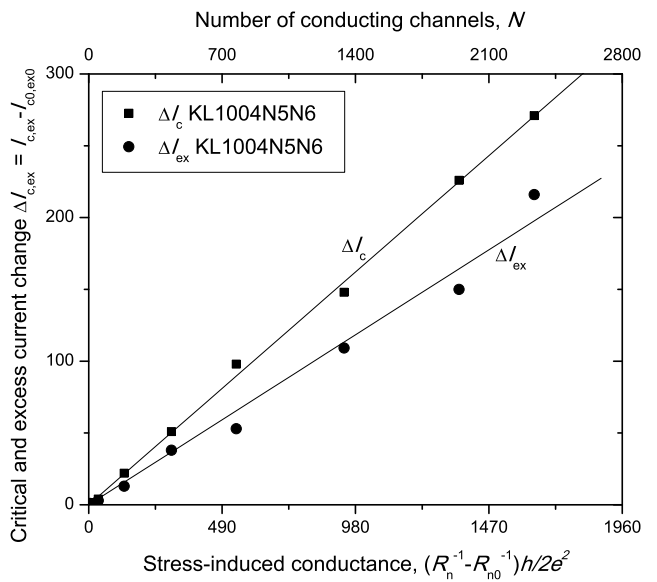}

\caption{\label{fig:sqpc-change-critical-excess}The change in critical current and
excess current caused by electric stressing as a function of the change in the
normal-state conductance (bottom scale) and the number of created channels (top scale) 
based on $\left\langle D\right\rangle =0.7$. The straight line is the linear fit giving
the average $I_{\mathrm{c}}$ per channel of 0.115 $\mu$A and $\delta I_{\mathrm{ex}}= 0.084$ $\mu$A/channel.}

\end{figure}

Since each SQPC carriers a supercurrent, the increase of the critical
current in the stressed junctions $\Delta I_{\mathrm{c}}=I_{\mathrm{c}}-I_{\mathrm{c}0}$ 
should be proportional to the number of conducting channels. This is shown
in Fig. \ref{fig:sqpc-change-critical-excess}. Using the assumption of equal channel
transparency and $\left\langle D\right\rangle \sim0.7$, we can estimate
the average contribution of each created channel $\delta I_{\mathrm{c}}=\Delta I_{\mathrm{c}}/N$
from Fig. \ref{fig:sqpc-fit-NvsV} to be about $0.115$ $\mu$A/channel. Using the experimental value
of $\Delta/e=1.35$mV and $\left\langle D\right\rangle =0.7$, from Eq. (\ref{eq:josephson-current})
we obtain the theoretical value of $\delta I_{\mathrm{c}}=0.147$ $\mu$A/channel,
a very close value.

The theory \cite{Averin1995} also explains the existence of the
excess current in the \IV\ curves as a nonzero average of the Josephson
ac current at finite voltages. It can be seen in Fig. \ref{fig:averin-comparison}
that the part of experimental \IV\ curves at $V>V_{\mathrm{g}}$ (region of
the excess current) agrees very well with the theory. The excess current
defined as $\Delta I_{\mathrm{ex}}=I_{\mathrm{ex}}-I_{\mathrm{ex0}}$ is
shown in Fig. \ref{fig:sqpc-change-critical-excess}, where $I_{\mathrm{ex}}$ was determined
by fitting the \IV\ curves at $V>V_{\mathrm{g}}$ to $I(V)=I_{\mathrm{ex}}+V/R_{\mathrm{n}}$
and index 0 identifies the initial, unstressed junction. The straight-line fit yields $\delta I_{\mathrm{ex}}= 0.084$ $\mu$A/channel, or $\delta I_{\mathrm{ex}}/I_{\mathrm{c}} = 0.73$ in very good agreement with the theory,\cite{Averin1995} giving $\delta I_{\mathrm{ex}}/I_{\mathrm{c}} = 2/\pi$ at $D\sim1$. 

So, by analyzing the additional subgap conductance, additional normal
state conductance, and additional Josephson currents induced by electric
stressing of the junction's barrier we obtained a self-consistent
picture. In this picture, after a point of soft breakdown has been
reached, atomic-size conduction channels with similar transparencies
are being formed in the barrier as the stressing progresses. We also 
proposed a scenario in which the formation of these channels is
a result of electric-field-stimulated ion migration away from the
barrier. Let us see if this scenario also agrees with the available experimental
data.

The treatment of ion transport in an electric field (electromigration)
is very well known and goes back to the original Frenkel defect theory
and works by Mott,\cite{Mott1947} and Cabrera and Mott.\cite{Cabrera1949}
Following \cite{Cabrera1949}, the ionic current density can be written
as
\begin{eqnarray}
J_{\mathrm{i}} & = & 2an\nu q\left[e^{-(W-qaE)/k_{\mathrm{B}}T}-e^{-(W+qaE)/k_{\mathrm{B}}T}\right]\nonumber \\
 & = & 2an\nu e^{-W/k_{\mathrm{B}}T}\sinh(qaE/k_{\mathrm{B}}T)\label{eq:ion-current}
\end{eqnarray}
where $n$ is the concentration of the mobile ions, $\nu$ is the
attempt frequency, $W$ is the activation energy, $q$ is the ion
charge and $a$ is the activation or half-jump distance.
Ionic (atomic) transport in solids is usually proceeds via vacancy
exchange mechanism (in which case $W$ is the height of the energy
barrier the ion needs to overcome) or via formation of Frenkel defects
(vacancy-interstitial pairs) in which case $W$ is the activation
energy for formation of a Frenkel defect. The total number of displaced
ions in the barrier can be hence presented as

\begin{equation}
N=At\sinh(qAE/k_{\mathrm{B}}T)=At\sinh(V/V_{0})\label{eq:sinh}\end{equation}
 where $t$ is the stress duration, $V$ is the voltage across the
barrier, $V_{0}=k_{\mathrm{B}}Td/qa$, and $d$ is the barrier thickness. 

Since in our model each displaced ion represents a new conduction
channel, in Fig. \ref{fig:sqpc-fit-NvsV} we plotted the ``effective''
number of conduction channels determined from the increase in the
junction conductance $N\cdot\left\langle D\right\rangle $= $G_{\mathrm{SQPC}}$$h/2e^{2}$
as a function of maximum voltage across the junction which develops
during each stressing at room temperature. The dependence given by Eq. (\ref{eq:sinh})
is also shown and fits the data very well, giving $V_{0}=0.045$$\pm0.005$
V. Assuming that oxygen ions are the mobile species, we can take $q=2e$,
$a=2.8$\AA\ (nearest-neighbor distance), $d=1$ nm, and $T=300$
K, and get from Eq. (\ref{eq:sinh}) $V_{0}=0.046$ V in excellent agreement
with the experimental data in Fig. \ref{fig:sqpc-fit-NvsV}. If instead
we use the value $qa=4.28\: e$\AA\ determined from direct measurements
of ionic current in alumina films,\cite{Dignam1968} the fit to the
experimental data in Fig. \ref{fig:sqpc-fit-NvsV} gives a tunnel barrier
thickness $d=0.76$ nm which is in a better agreement with the barrier
thickness data \cite{Tolpygo2003,Cimpoiasu2004} obtained from differential
conductance vs voltage dependences.

\begin{figure}
\includegraphics{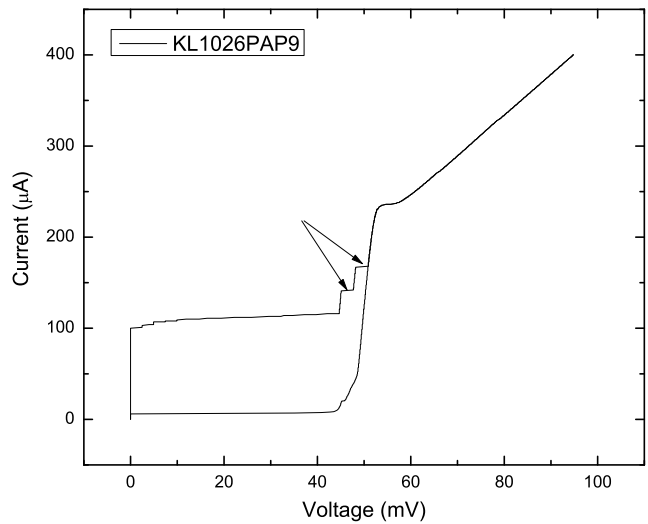}

\caption{\label{fig:array-IV} \IV\ curve of an as-fabricated 20-junction series
array. Two junctions with significantly higher $I_{\mathrm{c}}$ 
than the rest of the array are shown by arrows. Whereas the characteristics of these two
junctions are qualitatively similar to those of the electrically-stressed junctions
in this study, the amount of electric current that is required to
damage the stressed junctions ($I_{\mathrm{SB}}\, \times $(no. of junctions)) can not be supplied by 
the plasma processes employed during the fabrication. Hence, this fabrication-induced variation
in $I_{\mathrm{c}}$\ in this array is unlikely to be caused by electron current-induced breakdown
of the tunnel barriers.}

\end{figure}
A secondary objective of this study was to relate the studied above
effect of dc electrical stress with anomalous \IV\ characteristics
of Josephson junction arrays frequently observed for many fabrication
processes \cite{TolpygoISEC2007,TolpygoEUCAS2007,Hidaka2008}. As
was described in Sec. \ref{sec:INTRODUCTION}, usually one or two
junctions at the ends of the array demonstrate an $I_{\mathrm{c}}$ that is
significantly larger than the average $I_{\mathrm{c}}$ of the other junctions
in the array, see Fig. \ref{fig:array-IV}. Clearly two junctions
have an $I_{\mathrm{c}}$ that is significantly larger than the average $I_{\mathrm{c}}$ of
the rest of the junctions. Usually, the junction with the highest
deviation from the average is the last junction in the array, the
one which has the base electrode directly connected to the ground
plane layer.\cite{TolpygoISEC2007,TolpygoEUCAS2007} The second deviating
junction is the first junctions in the array, the one which is connected
by the wiring layer to the chip contact pads.\cite{TolpygoISEC2007}
These usually large deviations in $I_{\mathrm{c}}$ are also reflected in the so-called
knee region of the \IV\ curve where the knee current of the junctions
in question is much higher and can be distinctly seen in the normal
branch of the curve. Also the subgap conductance corresponding to
these two junctions is significantly higher than the average. MAR
steps in the subgap region of these junctions have also been observed.\cite{TolpygoEUCAS2007}

Qualitatively, these anomalous junctions look like the junctions subjected
to electric stress as described above. Electric currents can flow
through some of the Josephson junctions during their fabrication,
e.g., during reactive ion etching. The question is whether the amplitude
of the current is high enough to cause degradation of the barrier
in many junctions on the wafer. From the results of dc electric stressing,
we have found that the typical breakdown current in the smallest junctions
used in superconductor integrated circuits is $I_{\mathrm{SB}}^{-}\sim32$
mA at which a critical voltage develops across the barrier, i.e.,
$I_{\mathrm{SB}}/I_{\mathrm{c}}\sim250$. The typical 0.25-cm$^{2}$ test chip
contains $\sim20$ Josephson junction arrays. Therefore, to cause
breakdown of all 20 grounded junctions in these arrays, it would need
a current density of $\sim2.6$ A/cm$^{2}$ because for plasma-induced
current the grounded junctions are connected in parallel. There are
also several test chips spread out on the 150-mm wafers. So the total
current through the wafer should be huge. The typical rf power used
for etching in our process is from 40 to 150 W and peak-to-peak voltage
is from 305 V to 900 V depending on the circuit layer, corresponding
to power density from 88 mW/cm$^{2}$ to 332 mW/cm$^{2}$. Although,
rf power is coupled capacitively to the wafer and to the circuits'
ground plane, the rf current can flow directly through the junctions
which are galvanically coupled to the ground plane. Estimating the
amplitude of electric current supplied by the rf source, we find it
to be in the range of 0.5 A to 0.7 A, corresponding to the current
density from 1 mA/cm$^{2}$ to 1.55 mA/cm$^{2}$. This is clearly
too small a current to cause significant damage to oxide barriers
in Josephson junctions. Therefore, we conclude that, although exposure
to processing plasma has a potential in creating voltage differences
on the wafer surface which are large in comparison with the typical
breakdown voltages in ultrathin oxide barriers,\cite{Cheung2000}
the typical etching plasmas used in superconductor integrated circuits
fabrication do not supply enough current to electrically break down
the tunnel barriers of Josephson junctions due to their high tunneling
conductance. The same conclusion was also reached in \cite{TolpygoEUCAS2007}.

Because of these results, it is likely that the damage to junctions
galvanically coupled to the ground plane or other circuit layers observed
in \cite{TolpygoISEC2007} is chemical or electrochemical in nature.
For instance, it can be due to diffusion or electromigration of impurity
atoms from the layers in direct contact with the junction to the junction's
electrodes and to the barrier, e.g., hydrogen atoms dissolved in Nb.
In many respects the features of oxide barrier damage by H$^{+}$
electromigration should be similar to that of electron  current damage.
However, the required currents can be substantially lower because
hydrogen is known to be highly mobile in Nb. Therefore, it should
be much easier to cause diffusion of hydrogen atoms in Nb than
to displace oxygen atoms from the \alox\ oxide barrier. This damage
mechanism will be considered separately.

\section{CONCLUSION\label{sec:CONCLUSION}}

We have studied the effect of dc electric stress (current) applied
at room temperature to \nbaloxnb junctions with $J_{\mathrm{c}}$
= 1.0 and 4.5 \kacm\ on their Josephson and quasiparticle tunneling
properties. We have observed a very soft breakdown: above a certain
threshold stress current $I_{\mathrm{SB}}$ the changes in \IV\ characteristics
indicate formation in the barrier of additional conduction channels
with increased transparency. These channels reveal themselves in dramatic
increase in the subgap conduction, appearance of subharmonic current
steps (multiple Andreev reflections), an increase in the critical
current, a decrease in the normal state resistance, and an increase
in the $I_{\mathrm{c}}R_{\mathrm{n}}$ product. The range of the stress
currents where these gradual changes occur is quite broad, from $I_{\mathrm{SB}}$
to $\gtrsim 2I_{\mathrm{SB}}$, corresponding to $\sim 250I_{\mathrm{c}}$ to
$\gtrsim 500I_{\mathrm{c}}$. The threshold (breakdown) current depends on the stress
current polarity and is larger when the Nb counter electrode is at
a positive potential.

We have suggested that the soft breakdown proceeds via formation of defects in the barrier
as a result of ion migration in the applied electric field. The stress polarity
dependence was explained as due to internal electric field caused by the difference in the work 
functions of the junction electrodes. The observed breakdown fields $E_{\mathrm{c}} \sim 3.85 \times 10^8$ V/m ($V_{\mathrm{c}} \cong 0.385$ V 
across the barrier) correspond to the magnitude required for near-neighbor displacement of oxygen ions.
 
From the changes in the tunneling characteristics in the superconducting
state we concluded that the total area of the additional conduction
channels is much smaller than the area of the initial tunnel barrier,
and that they can be considered as connected in parallel to the initial
tunnel junction whose properties remain virtually unchanged by electric
stressing. Using this parallel connection model, we estimated the
transparency of the additional conduction channels by comparing their
\IV\ characteristics to the nonstationary theory of current transport
in superconducting quantum point contacts by Averin and Bardas. We
found that all the channels have similar transparencies $D$ in the range
from $\sim0.6$ to $\sim0.8$. We found an excellent agreement
of the shape of \IV\ characteristics of the conduction channels created
by electric stressing with the Averin-Bardas theory for SQPCs.\cite{Averin1995}
From the change in the normal-state resistance we calculated the effective
number of the additional conduction channels $N$. The observed changes
in the junction critical current per channel and the excess current
per channel were found to be consistent with the predictions of the microscopic
theory for SQPCs.

We also found an exponential dependence of the number of additional
conduction channels created by electric stressing on the stress voltage,
$N\sim\sinh(V/V_{0})$. The characteristic voltage $V_{0}$ was found
to be in excellent agreement with the proposed model of defect formation
by ion electromigration out of the barrier.

Due to the high tunneling conductance of oxide barriers used in superconductor
digital electronics it is unlikely that breakdown currents can be
reached during plasma processing steps of junction fabrication. However,
it is possible that electromigration and diffusion of impurity atoms
in Nb layers is responsible for pattern-dependent tunnel barrier degradation
frequently observed in superconductor integrated circuits.

\begin{acknowledgments}
We would like to thank Daniel Yohannes, Richard Hunt, and John Vivalda
for their part in wafer processing. Many discussions with Alex Kirichenko,
Timur Filippov, Vasili Semenov and Dmitri Averin are highly appreciated.
We would also like to thank D. Averin for providing the program for calculating 
the theoretical curves shown in Fig. \ref{fig:averin-comparison}. We are also grateful to Deborah Van Vechten for her interest and
support of this research. This work was supported by ONR Grants N000140810224 and N000140710093. 
\end{acknowledgments}

\end{document}